\documentclass[11pt]{article}
\usepackage[utf8]{inputenc}

\usepackage{a4}
\usepackage{amsmath}
\usepackage{amsthm}
\usepackage{amssymb} 
\usepackage{euler} 
\usepackage[round]{natbib}
\usepackage[hang,flushmargin]{footmisc}

\newcommand{\X}{\mathcal{X}}
\newcommand{\W}{\mathcal{W}}
\def\O{\mathcal{O}}
\newcommand{\ap}{{\sss\rm ap}}
\newcommand{\g}{\sss\rm stat}
\newcommand{\ret}{{\sss\rm ret}}
\newcommand{\av}{{\sss\rm av}}
\def\max{{\sss\rm max}}
\def\min{{\sss\rm min}}
\newcommand{\up}{\uparrow}
\newcommand{\down}{\downarrow}
\newcommand{\Wop}{\hbox{$\displaystyle{\mathop{\W}^{\circ}}$}}
\font\tenssbf=cmssdc10

\newcommand{\Ev}{{\tenssbf E}}
\newcommand{\V}{{\tenssbf V}}
\def\S{{\tenssbf S}}
\newcommand{\sss}{\scriptscriptstyle}
\newcommand{\reals}{\mathbb{R}}

\newtheorem{theorem}{Theorem}
\newtheorem{consequence}[theorem]{Consequence}
\newtheorem{statement}[theorem]{Statement}

\begin{document}
\tolerance 10000      

\title{
On the statistical viewpoint concerning\\ 
the second law of thermodynamics \\
\textit{\Large- a reminder on the Ehrenfests' urn model -}}

\author{
Domenico Giulini\\
ZARM Bremen\\
Institute for Theoretical Physics\\
Leibniz University Hannover\\
\texttt{giulini@itp.uni-hannover.de}
}      

\date{}
\maketitle

\begin{abstract}\noindent 
In statistical thermodynamics the 2nd law 
is properly spelled out in terms of 
\emph{conditioned} probabilities.  
As such it  makes the statement that 
``entropy increases 
with time'' \emph{without} preferring a 
time direction. In this paper we wish to 
explain and illustrate this statement
in terms of the Ehrenfests' urn model 
in a way that  hopefully adds some 
clarifying aspects concerning the role of
time-conditioned probabilities. We will 
relate past- and future-conditioned 
probabilities through Bayes' rule, which 
allows us to explicitly state what is 
meant by time-reversal invariance in this 
context. 

This paper is my contribution to 
the book \emph{From Quantum to Classical -- 
Essays in Honour of H.-Dieter Zeh}, edited by Claus Kiefer, that appeared in 2022 as volume 204
in the series \emph{Fundamental Theories of Physics} at Springer Verlag.
\end{abstract}

 \section{Prologue}
This contribution is dedicated to the 
memory of H. Dieter Zeh, with whom I had
many discussions over a period of time 
that easily spans a quarter of a century.
These discussions were usually
controversial, sometimes \emph{very}
controversial, so that often we could 
only agree do disagree. From these discussions 
I learned a lot!

I came to know Zeh - the university teacher
- in my earlier student days.  In June 
1984 a friend gave me Zeh's 
``Die Physik der Zeitrichtung'' as a birthday 
present \citep{Zeh:PhysikDerZeitrichtung-1}. 
Below the preface my friend wrote in
his own hand: ``I am sure this book will
find your undivided approval'', and
it did! Not so much because I felt competent 
enough to judge the content, but because I
felt a degree of veracity behind it that appealed to me very much. 

The book of which I received a copy was 
the first edition, based on notes of lectures Zeh had given since 1979. When 
the first edition finally came out it 
quickly circulated amongst the younger generation of students, even outside 
the physics department (in fact, my 
friend was in the department of chemistry 
and knew about the book before I did). 
For many of us Zeh represented 
the serious and uncompromising urge for real 
``understanding'' that many of us hoped to
find at the university and, in particular,
in the department of theoretical physics.
This hope was not always fulfilled but 
Zeh was definitely someone proving that 
our hopes had not been in vain.%
\footnote{To give a contrasting example, 
I remember from my first lecture on 
quantum field theory, in which after the  
scheme of canonical quantisation was 
introduced and the interaction picture 
mentioned, the professor said: ``There 
is a theorem due to Rudolf Haag, 
according to which the interaction 
picture does not exist; but we shall 
henceforth ignore that!'' How should a 
serious beginning student deal with such 
a casually tossed comment?} 

I should add that our trust was based 
not so much on the fact that Zeh talked
about the ``big issues'', like ``arrow of 
time'', or ``interpretation of quantum
mechanics'', 
which clearly fascinate all beginning
student 
alike, but rather on the fact that he
touched upon these issues while at the same
 time striving for full clarity as regards
the ``small issues''. The only lecture I
took by Zeh was on analytical mechanics.
From that I remember his explanation of the 
Euler angles, which he gave by means of 
meticulous blackboard drawing that he had
prepared before the lecture, showing 
four systems of orthonormal frames in
different 
colours with the relevant rotation axes and 
angles. His comment was that he did not
understand the corresponding drawing in 
the standard textbook then widely used, 
so he developed everything from scratch
again. 

Many years later Zeh gave me the third
english 
edition \citep{Zeh:DirectionOfTime-3} as a 
present. That edition contains a new Appendix
on a simple numerical toy model, the so called 
\emph{ring model}, taken from 
\citet[chapter\,3]{Kac:1959}, which is 
meant to illustrate the concept of a 
``Zwanzig projection'', that plays a most central 
role in Zeh's book, as one can already see 
from the book's bibliography, that contains 
43 entries for that term (the only entry under 
``Z''). Zeh also mentions the Ehrenfest's 
\emph{urn-model} as a popular and widely 
known one to illustrate conceptual points 
connected with statistical statements in physics. 
I remember discussing that with Zeh and that 
I praised the Ehrenfests' model for its ability 
to illustrate basic but non-trivial concepts 
from statistical mechanics by means of 
\emph{exact} mathematical expressions. 
It was then that I worked out some details 
of that model, just for my own understanding,  
illustrating -- as Zeh used to say -- the 
``fact-like'' rather than ``law-like'' nature 
of entropy increase. This is what I wish to 
present here.  

Following the third, there were two more editions 
of Zeh's classic text, the last (fifth) in Springer's 
\emph{Froniers Collection} \citep{Zeh:DirectionOfTime-5}, that I reviewed in \citet{Giulini:2008}.

\section{Introduction}
The point that the likely statistical increase 
of entropy does  \emph{not} as such prefer a 
direction of time has been first made by 
\citet{Ehrenfests:1907} in connection
with their \emph{urn model} \citep{Ehrenfests:1906}; see also their
general review \citet{Ehrenfests:1912}
(english translation \citet{Ehrenfests:1990}).
It has again been emphasised by 
\citet{Weizsaecker:1939}\footnote{The conclusions 
Weizs\"acker drew from this insight are, 
however, problematic; see,
\citet{Kiefer:2014}}.
This insight is not new and should 
be a commonplace, though that is far 
from true according to my experience. 

Without going into any analytical 
details, Zeh said it  very clearly 
at the beginning of paragraph\,3 of 
\citet[p\,40]{Zeh:DirectionOfTime-5} on the 
thermodynamical arrow of time: 
\begin{quote}
``While statistical considerations are 
indeed essential for the understanding 
of thermodynamical concepts, statistics 
as a method of counting has nothing a 
priori to do with dynamics. Therefore, 
it cannot by itself explain dynamically 
`irreversible' processes -- characterised 
by $\{\mathrm{d}S/\mathrm{d}t\}_{\mathrm{int}}>0$.
This requires \emph{additional} assumptions, 
which often remain unnoticed, since they 
appear `natural' to our prejudiced way 
of thinking in terms of causes (exclusively
in the past). These hidden assumptions 
have therefore to be carefully investigated 
in order to reveal the true origin of 
the thermodynamical arrow.''%
\footnote{In the first (german) edition 
and in the following english editions up 
to, and including, the third, the  ``has 
nothing a priori to do with dynamics'' 
reads instead: ``has nothing a priori to 
do with the physical concept of time or 
its direction'' \citet[p\,37]{Zeh:DirectionOfTime-3}, 
or ``...jedoch hat die Statistik als 
mathematische Disziplin nichts mit der 
physikalischen Zeit zu tun und vermag daher den 
Zeitpfeil auch nicht zu begr\"unden''
\citep[p\,23]{Zeh:PhysikDerZeitrichtung-1}.}.
\end{quote}
Zeh continues by comparing the four 
possibilities of processes in time
\[
\text{probable / improbale state}\quad
\mathop{\longrightarrow}^{t}\quad 
\text{probable / improbable state}
\]
pointing out, in particular, that there are as many 
processes from improbable to probable than from 
probable to improbable states. Hence, an asymmetry 
in their number of occurrences must be connected 
with an \emph{additional}, symmetry breaking 
assumption of an improbable state at one end of 
the time axis. Zeh calls that prescribed state 
the ``initial'' one, which at this point may be 
read as an implicit definition of ``initial'', namely 
as that end at which the condition is put, for 
otherwise there is so far no objective difference 
between the two ends of the time axis. 

In this contribution I will employ the 
celebrated urn model to exemplify these points. 
This model has once even been called 
``probably one of the most instructive models 
in the whole of physics''\citet[p\,73]{Kac:1959}. 
The rather simple analytical features of this 
model help to guide one's own expectations and 
reduce the danger of possible misconceptions. 
For the issue to be discussed here, namely 
that of the ``likely increase'' of entropy
is a time symmetric statement, possible
 misconceptions have to do with 
a failure to appreciate the fact that the 
probabilities are conditioned in time,
and that their interpretation, namely as being 
either ``retarded'' or ``advanced'', is 
in itself indistinguishable unless a time 
orientation has already been established.

I will proceed as follows: In the next section 
I will try to put the qualitative statement 
just made into more precise words. In the 
following section this is then quantitatively 
analysed in terms of the urn model, where 
probabilities can actually be calculated in an 
explicit and elementary way. The remaining 
sections then discuss Boltzmann entropy, Gibbs 
entropy and $H$-theorem, and the thermodynamic 
limit and deterministic dynamics. 
Some elementary notions from probability
theory that we made freely use 
of are collected in a short appendix.

\section{The basic statements}
In this section we start by making some 
of the previous statements more precise.
We think of an idealised system, whose 
state may only change at sharp,  discrete 
times. This allows us to speak  unambiguously 
about ``next'' and ``previous'' points in 
time. Now we make the following 
\medskip

\noindent
\textbf{Assumption.}
At time $t_i$ the system is in a state $z(t_i)$ 
of \emph{non-maximal} entropy. The statistical 
2nd law now makes the following statement about 
conditioned probabilities (the condition, which 
is just this assumption, will not be repeated):  

\begin{statement}
The probability, that the state $z(t_i)$ will develop 
\emph{in the future} to a state $z(t_{i+1})$ of larger entropy,
is larger than the probability for a development into a state 
of smaller entropy. 
\end{statement}

\begin{statement}
The probability, that the state $z(t_i)$ has developed in 
the \emph{past} from a state $z(t_{i-1})$ of larger entropy, 
is larger than the probability of a development from a state 
of smaller entropy. 
\end{statement}

\begin{consequence}
The likely increase of entropy in the 
future state-development $z(t_i)\mapsto z(t_{i+1})$ does not imply 
a likely decrease for the (fictitious) past development
$z(t_i)\mapsto z(t_{i-1})$, but rather 
also a likely increase. 
\end{consequence}

\begin{consequence}
The most likely development $z(t_{i-1})\mapsto z(t_i)$
is that of decreasing entropy. Somewhat ironically, one may 
say that it is more likely for the state $z(t_i)$ to come 
about through the improbable development from a more probable 
state $z(t_{i-1})$ than through the 
probable development from an improbable state. 
\end{consequence}

To properly understand the last consequence, recall that 
our \textbf{condition} is placed on $z(t_i)$, that is at time 
$t_i$.  For $z(t_i)\mapsto z(t_{i+1})$ this means a 
\emph{retarded} or \emph{initial} condition, for 
$z(t_{i-1})\mapsto z(t_i)$, however, an  \emph{advanced} or 
\emph{final} condition. It is this change 
of condition which makes this behaviour of entropy 
possible. 

\begin{consequence}
The mere (likely) increase of entropy does not provide 
an orientation of time. It does \emph{not} serve to 
define a `thermodynamic arrow of time'. Rather, an 
orientation is usually given by considering a definite 
time-interval (usually of finite length) and imposing 
a low-entropy condition at one of the two \emph{ends} 
of that interval. Without further structural elements 
that would serve to distinguish the two ends, the 
apparently existing  \emph{two} possibilities to place 
the low-entropy conditions are, in fact, identical. 
An apparent distinction is sometimes introduced by 
stating that the condition at one end is to be understood 
as \emph{initial}. But at this level this merely 
defines \emph{initial} to be used 
for that very end at which the condition 
is placed. 
\end{consequence}
\goodbreak

\section{The Urn-Model}
This model was introduced by 
\citet{Ehrenfests:1906,Ehrenfests:1907}
and quickly entered textbooks and other 
pedagogical oriented discussions. More 
detailed mathematical discussion of it 
are contained in \citet{Kohlrausch.Schroedinger:1926}, 
reprinted in \citet[pp.\,349-357]{Schroedinger:GA1} and \citet{Kac:1947,Kac:1959}. \citet{Kohlrausch.Schroedinger:1926}
also report on actual experiments done in 
order to determine the Boltzmann $H$-curve for this model. 
  
Think of two urns,  $U_0$ and $U_1$, among which one distributes 
$N$ numbered balls. For exact equipartition to be possible we 
assume $N$ to be even. A \emph{microstate} is given by the individual numbers (names) 
of balls contained in $U_1$. (The complementary set of numbers 
then label the balls in $U_0$.) To formalise this, we associate a 
two-valued quantity $x_i\in\{0,1\}$, $i\in\{1,\dots,N\}$, to 
each ball, where $x_i=0$ ($x_i=1$) stands for the $i$'th ball 
being in $U_0$ ($U_1$). This identifies the set of microstates,
which we will call $\Gamma$ (it corresponds to phase space), 
with $\Gamma=\{0,1\}^N$, a discrete space of of $2^N$ elements. 
It can be further identified with the set of all functions 
$\{1,\cdots,N\}\rightarrow\{0,1\}$, $i\mapsto x_i$. Mathematically 
speaking, the space $\Gamma$ carries a natural measure, 
$\mu_{\sss \Gamma}$, given by associating to each subset 
$\Lambda\subset\Gamma$ its cardinality: 
$\mu_{\sss \Gamma}(\Lambda)=\vert\Lambda\vert$. 
We now make the physical assumption, that the probability measure
(normalized measure) $\nu_{\sss \Gamma}:=2^{-N}\mu_{\sss \Gamma}$
gives the correct \emph{physical} probabilities. Note that this is a 
statement about the dynamics, which here my be expressed by saying, 
that in the course of the dynamics of the system, all microstates 
are reached equally often on time average. 

Physical observables correspond to functions  
$\Gamma\rightarrow\reals$. We call the set of such functions 
$\O$. Conversely, it is generally impossible to associate
a physically realisable observable to any element in $\O$.
Let $\{O_1,\dots,O_n\}=:\O_{\rm re}\subset\O$ be
the physically realisable 
ones\footnote{The subscript `re' 
can be read as abbreviation for `realised' or `relevant'.},
which we can combine into a single 
$n$-component observable
$O_{\rm re}\in\O^n$. If $O_{\rm re}:\Gamma\rightarrow\reals^n$
is injective, the state is determined by the value of $O_{\rm re}$.
In case of thermodynamical systems it is essential to be far 
away from injectivity, in the sense that a given value 
$\alpha\in\reals^n$ should have a sufficiently large pre-image 
$O_{\rm re}^{-1}(\alpha)\subset\Gamma$. The coarse-grained 
of macroscopic state space is then given by the image
$\Omega\subset\reals^n$ of the realized observables $O_{\rm re}$.
To every macrostate $\alpha\in\Omega$ corresponds a set of 
microstates: $\Gamma_{\alpha}:=O_{\rm re}^{-1}(\alpha)\subset\Gamma$.
The latter form a partition of  $\Gamma$: 
$\Gamma_{\alpha}\cap\Gamma_{\beta}=\emptyset$ if  
$\alpha\not =\beta$ and 
$\bigcup_{\alpha\in\Omega}\Gamma_{\alpha}=\Gamma$. 
 
The realised observable for the urn-model is given by the number 
of balls in $U_1$, that is, $O_{\rm re}=\sum_{i=1}^N x_i$.
Its range is the set $\Omega=\{0,1,\dots,N\}$ of macrostates, which 
contains $N+1$ elements. The macrostates are denoted by $z$. 
To $z$ there corresponds the set $\Gamma_z$ of 
$\left(N\atop z\right)$ microstates. The probability measure 
$\nu_{\sss \Gamma}$ induces so-called 
`a-priori-probabilities' for 
macrostates $z$:     
\begin{equation}
W_{\ap}(z)=\nu_{\null_{\Gamma}}(\Gamma_z)=2^{-N}\left(N\atop z\right).
\label{1}
\end{equation}
 
Let$X:\Omega\rightarrow\reals$ be the random variable 
$z\mapsto X(z)=z$. Its expectation value, 
denoted by $\Ev$, and its standard deviation,
denoted by $\S$,  with respect to the 
a-priori-distribution~(\ref{1}) are given 
by
\begin{eqnarray}
\Ev(X,\hbox{ap})
&=& \frac{N}{2}, \label{2}\\
\S(X,\hbox{ap})
&=&\frac{\sqrt{N}}{2}.\label{3}
\end{eqnarray}
This follows from 
\begin{equation}
\label{F1}
\begin{split}
E(X;\hbox{ap})
&=2^{-N}\sum_{z=1}^N z\,\left(N\atop z\right)
 =2^{-N}N\sum_{m=0}^{N-1}\left(N-1\atop m\right)=\frac{N}{2}\,,\\
E(X^2-X;\hbox{ap})
&=2^{-N}\sum_{z=2}^N z(z-1)\,\left(N\atop z\right)\\
&=2^{-N}N(N-1)\sum_{m=0}^{N-2}\left(N-2\atop m\right)
 =\frac{N(N-1)}{4}\,,\\
S^2(X;\hbox{ap})
&=E(X^2-X;\hbox{ap})+E(X;\hbox{ap})-E^2(X;\hbox{ap})\,.
\end{split}
\end{equation}

The system has a Markovian random evolution, which is defined 
as follows: At every time $t_i$, where 
$i=\{0,1,2,\cdots\}$ with $t_j>t_i$ for $j>i$, a random 
generator picks a number $n$ in the interval $1\leq n\leq N$.
Subsequently the ball with number $n$ changes the urn. There are 
two possibilities: The ball with number $n$ has been in urn 
$U_0$ so that the change of macrostate is given by 
$z\rightarrow z+1$. Alternatively, the ball has been in $U_1$ 
and the change of macrostate is given by $z\rightarrow z-1$. 
The conditional probabilities, $W(z\pm 1;t_{i+1}\vert z;t_i)$,
that given the state $z$ at time $t_i$ the evolution will yield 
the state $z\pm 1$ at time $t_{i+1}$, are given by 
\begin{alignat}{3}
& W(z+1;t_{i+1}\vert z;t_i) &&\,=\,\frac{N-z}{N}
&&\,=\,:\, W_{\ret}(z+1\vert z), \label{4a}\\
& W(z-1;t_{i+1}\vert z;t_i) &&\,=\quad \frac{z}{N}
&&\,=\,:\, W_{\ret}(z-1\vert z). \label{4b}
\end{alignat}
Since these are independent of time, we can suppress the 
arguments $t_i$. We just have to keep in mind that the left 
entry, $z\pm 1$, is one time step \emph{after} the time of 
 $z$, that is, the probabilities are \emph{past-conditioned} or \emph{retarded}. 
We indicate this by writing $W_{\ret}$.  

Let $W(z;t_i)$ denote some chosen absolute probability for the 
state to be $z$ at time $t_i$ and $W_i:z\rightarrow W(z;t_i)$
the probability distribution at time $t_i$. The dynamics described 
above will now induce a dynamical law, $W_i\rightarrow W_{i+1}$,
on such distributions, given by 
\begin{eqnarray}
W(z;t_{i+1}) &=& W(z;t_{i+1}\vert z+1;t_i)\, W(z+1;t_i)\nonumber\\
             &+& W(z;t_{i+1}\vert z-1;t_i)\, W(z-1;t_i)\label{5}\\
             &=& \frac{z+1}{N}\,   W(z+1;t_i) 
              + \frac{N-z+1}{N}\, W(z-1;t_i),
\label{6}
\end{eqnarray}
whose Markovian character is obvious. To be sure, 
$W_i$, $i>0$, will depend on the initial distribution $W_0$. 
This dependence will be essential if $W_0$ is far from equilibrium 
and the number of time steps $i$ not much larger than the number 
$N$ of balls. Conversely, one expects that for $W_i$ will 
approach an equilibrium distribution $W_{\g}$ for $i\gg N$, 
where $W_{\g}$ is independent of $W_0$. Its uniqueness is shown 
by 
\begin{theorem}
A distribution $W_{\g}$ which is stationary under (\ref{6}) 
is uniquely given by $W_{\ap}$ in (\ref{1}).
\end{theorem}

\begin{proof}
We show, that $W_{\g}$ can be uniquely determined from (\ref{6}).
To this end, we assume a time independent distribution 
$W_{\g}$ and write (\ref{6}) in the form
\begin{equation}
W_{\g}(z+1) = \frac{N}{z+1}\, W_{\g}(z)
            - \frac{N-z+1}{z+1}\, W_{\g}(z-1).
\label{7}
\end{equation}
Since $W_{\g}(-1)=0$ we have for $z=0$ that $W_{\g}(1)=NW_{\g}(0)$,
hence recursively $W_{\g}(2)=\frac{1}{2}N(N-1)W_{\g}(0)$ 
and $W_{\g}(3)=\frac{1}{6}N(N-1)(N-2)W_{\g}(0)$. By induction we get 
the general formula $W_{\g}(z)=\left(N\atop z\right)W_{\g}(0)$.
Indeed, inserting this expression for $z$ and $z-1$ into the right 
hand side of $(7)$, we obtain 
\begin{eqnarray}
W_{\g}(z+1) &=& \left[\frac{N}{z+1}\left(N\atop z\right)
  -\frac{N-z+1}{z+1}\left(N\atop z-1\right)\right]\,W_{\g}(0)
\nonumber\\
&=& (N-z)\frac{N(N-1)\cdots (N-z+1)}{(z+1)!}\ W_{\g}(0)
\nonumber\\
&=& \left(N\atop z+1\right)\,W_{\g}(0).
\label{8}
\end{eqnarray}
The value of $W_{\g}(0)$ is finally determined by the normalization 
condition:
 \begin{equation}
1=\sum_{z=0}^{n} W_{\g}(z)=W_{\g}(0)\sum_{z=0}^N\left(N\atop z\right)
 =W_{\g}(0)\ 2^N\Rightarrow W_{\g}(0)=2^{-N}.
\label{9}
\end{equation}
\end{proof}

\subsection{Future-conditioned probabilities and Bayes' rule}
Consider a probability space and a set of events, 
$\{A_1,\dots,A_n\}$,
which is 1.)~complete, i.e. $A_1\cup\cdots\cup A_n=\mathbf{1}$
(here $\mathbf{1}$ denotes the certain event), and 2.)~mutually 
exclusive, i.e. $i\not =j\Rightarrow A_i\cap A_j=\mathbf{0}$
(here $\mathbf{0}$ denotes the impossible event). The probability 
of an event $B$ then obeys the well known rule
$W(B)=\sum_{k=1}^n W(B\vert A_k)W(A_k)$. This is just what we used 
in (\ref{5}). Now, \emph{Bayes' rule},
which we here regard as an 
independent assumption\footnote{Therefore 
we avoid to call it Bayes' theorem.},
will now allow us to deduce the 
inversely conditioned probabilities: 
\begin{equation}
W(A_k\vert B)=\frac{W(B\vert A_k)W(A_k)}{
  \sum_{i=1}^n W(B\vert A_i)W(A_i)}.
\label{10}
\end{equation}
We now identify the $A_i$ with the $N+1$ events $(z';t_i)$
at the fixed time $t_i$, where $z'=0,\dots,N$, and $A_k$
with the special event $(z\pm 1;t_i)$. Further we identify the 
event $B$ with $(z;t_{i+1})$, i.e. with the occurrence of $z$ 
at the \emph{later} time $t_{i+1}$. Then we obtain: 
\begin{eqnarray}
W(z\pm 1;t_i\mid z;t_{i+1})
&=&{W(z;t_{i+1}\vert z\pm 1;t_i)W(z\pm 1;t_i)\over
   \sum_{z'=0}^NW(z;t_{i+1}\vert z';t_i)W(z';t_i)}
\label{11}\\
&&\nonumber\\
&=&{W(z;t_{i+1}\vert z\pm 1;t_i)W(z\pm 1;t_i)\over
   W(z;t_{i+1})}.
\label{12}
\end{eqnarray}
Hence, given $W_i$, a formal application of Bayes' rule allows us to 
express the future conditioned (`advanced') probabilities in terms 
of the past conditioned (`retarded') ones. In our case we think of 
the latter ones as given by  (\ref{4a}-\ref{4b}). Hence we obtain 
the conditioned probability for $(z\pm 1;t_i)$, given that at the 
later time $t_{i+1}$ the state will $z$ occur:   
\begin{eqnarray}
W(z+1;t_i\vert z;t_{i+1})
&=&{W(z+1;t_i)\over W(z+1;t_i)+{N-z+1\over z+1}W(z-1;t_i)},
\label{13a}\\
W(z-1;t_i\vert z;t_{i+1})
&=&{W(z-1;t_i)\over W(z-1;t_i)+{z+1\over N-z+1}W(z+1;t_i)}.
\label{13b}
\end{eqnarray}  

\subsection{Flow equilibrium}
The condition for having flow equilibrium for the pair of times 
$t_i,t_{i+1}$ reads 
\begin{equation}
W(z\pm 1;t_{i+1}\vert z;t_i)W(z;t_i)
=W(z;t_{i+1}\vert z\pm 1;t_i)W(z\pm 1;t_i).
\label{14}
\end{equation}
It already implies $W_i=W_{\ap}$, since 
(\ref{4a}-\ref{4b}) give%
\footnote{
Without using (\ref{4a}-\ref{4b}) one gets
\begin{eqnarray}
  W(z\pm 1;t_{i+1}\vert z;t_i)W(z;t_i)
&=& W(z;t_{i+1}\vert z\pm 1;t_i)W(z\pm 1;t_i)\nonumber\\
&=& W(z\pm 1;t_i\vert z;t_{i+1})W(z;t_{i+1})\,,
\label{F3}
\end{eqnarray}
where the last equality is the identity 
$W(a\vert b)W(b)=W(b\vert a)W(a)$. The local (in time) 
condition of flow equilibrium is therefore equivalent to 
(cf.~\ref{15})
\begin{equation}
{W(z\pm 1;t_{i+1}\vert z;t_i)\over W(z\pm 1;t_i\vert z;t_{i+1})}
={W(z;t_{i+1})\over W(z;t_i)}.
\label{F4}
\end{equation}
}
 $W(z+1;t_i)={N-z\over z+1}W(z;t_i)$ which leads to 
$W(z;t_i)=\left(N\atop z\right)W(0;t_i)$. Since $1=\sum_zW(z;t_i)$
we have $W(0;t_i)=2^{-N}$. Using Theorem\,1, we conclude that 
flow equilibrium at $t_i$ implies $W_j=W_{\ap}$ for $j\geq i$. 

\subsection{Time-reversal invariance}
To be distinguished from flow equilibrium is time-reversal 
invariance. The latter is given by the following equality of 
past- and future-conditioned probabilities:  
\begin{eqnarray}
W(z\pm 1;t_{i+1}\vert z;t_i)
&&=\quad
W(z\pm 1;t_i\vert z;t_{i+1})
\label{15}\\
&&{\mathop{=}^{{\sss (\ref{12})}}}\quad W(z;t_{i+1}\vert z\pm 1;t_i)
{W(z\pm 1;t_i)\over W(z;t_{i+1})},
\label{16}\\
\mathop{\Longleftrightarrow}^{{\sss (\ref{4a},\ref{4b})}}\quad
W(z;t_{i+1}) &&=\quad {z+1\over N-z}\, W(z+1;t_i) \label{17a}\\
         &&=\quad {N-z+1\over z}\, W(z-1;t_i).   \label{17b}
\end{eqnarray}
It is interesting to note that the condition of time-reversal
invariance is weaker that that of flow equilibrium. The former 
is implied by, but does not itself imply,
the equilibrium 
distribution. Let us explain this in more detail: 
Equations (\ref{17a}-\ref{17b}) imply (\ref{6}), since 
${N-z\over N}\times(\ref{17a})+{z\over N}\times (\ref{17b})=(\ref{6})$.
Hence (\ref{17a}-\ref{17b}) are stable under time evolution (\ref{6}).
Conversely, (\ref{17a}-\ref{17b}) is implied by (\ref{6}) and the 
following equation, expressing the equality of the right hand sides 
of (\ref{17a}) and (\ref{17b}):
\begin{equation}
W(z+1;t_i)={N-z\over z+1}\,{N-z+1\over z}\, W(z-1;t_i).
\label{18}
\end{equation}
Indeed, eliminating  $W(z+1;t_i)$ in (\ref{6}) using  
(\ref{18}), one gets
\begin{equation}
W(z;t_{i+1})={N-z+1\over z}\,W(z-1;t_i)\mathop{=}^{(18)}
{z+1\over N-z}\,W(z+1;t_i),
\label{19}
\end{equation}
hence  (\ref{17a}-\ref{17b}). Time-reversal invariance for future 
times is therefore equivalent to the `constraint' (\ref{18})
for the initial condition. It allows for a one-parameter family of 
solutions, since  it determines $W_i$ for given $p:=W(0;t_i)$
and $q:=W(1;t_i)$. Indeed, in analogy to the proof of Theorem\,1 
one gets $W_i(z)=p\left({N\atop z}\right)$ for $z$ even and 
$W_i(z)={q\over N}\left({N\atop z}\right)$ for $z$ odd. Since 
$\sum_{z={\rm even}}\left({N\atop z}\right)=
\sum_{z={\rm odd}}\left({N\atop z}\right)=2^{N-1}$, the 
normalization condition leads to 
$1=2^{N-1}(p+{q\over N})\Rightarrow q=N(2^{-(N-1)}-p)$.
This shows that $p\in[0,2^{-(N-1)}]$ faithfully parameterizes all 
distributions obeying  (\ref{18}). One should note that solutions 
to (\ref{18}) are closed under convex sums. In this way one sees, 
that the obtained distributions are the convex sum 
$W_i=pW^e+(1-p)W^o$ of the `even' distribution, 
$W^e(z)=(1-(-1)^{z-1})2^{-N}\left(N\atop z\right)$ and `odd'
distribution, $W^o(z)=(1-(-1)^z)2^{-N}\left(N\atop z\right)$.
Solutions to (\ref{18}) form a closed interval within the simplex 
$\Delta^N$, which connects the point $W^e$ in the 
${N\over 2}$--sub-simplex $\Delta^{13\dots N-1}$ with the point 
$W^o$ on the $({N\over 2}+1)$--sub-simplex $\Delta^{24\cdots N}$. 
If we call this interval  $\Delta^*$, we have
\begin{theorem}
The set $\Delta^*\subset\W$ is invariant under time evolution. 
The future development using  $W(z;t_{i+1}\vert z';t_i)$ and 
the past development using $W(z;t_i\vert z';t_{i+1})$ 
coincide.\footnote{Explicitly one can see the preservation of 
(\ref{18}) under time evolution (\ref{6})  as follows: 
Given that the initial distribution $W_i$ satisfies (\ref{18}),
the development (\ref{6}) is equivalent to (\ref{17a}-\ref{17b}).
Hence 
\begin{eqnarray}
W(z-1;t_i)&=&{z\over N-z+1}\,   W(z;t_{i+1})   \label{F5}\\
W(z+1;t_i)&=&{z+2\over N-z-1}\, W(z+2;t_{i+1})\,, \label{F6}
\end{eqnarray}
which allows to rewrite (\ref{18}) for $W_i$ into (\ref{18}) 
for $W_{i+1}$.
}
\end{theorem}

It is of central importance to note that the past development
is, mathematically speaking, \emph{not} the inverse operation to 
the future development. The reason being precisely that such a 
change in the direction of development is linked with a change 
from retarded to advanced conditionings in the probabilities.    

\section{General Consequences}
In the following we want to restrict to the equilibrium condition. 
In this case the future-conditioned probabilities are independent 
of the $t_i$ and we can write
$W(z\pm 1;t_i\vert z;t_{i+1})=:W_{\av}(z\pm 1\vert z)$. Hence we have: 
\begin{alignat}{3}
& W_{\ret}(z+1\vert z)
&&\,=\,W_{\av}(z+1\vert z)
&&\,=\,{N-z\over N}\,,\label{20a}\\
& W_{\ret}(z-1\vert z)
&&\,=\,W_{\av}(z-1\vert z)
&&\,=\quad {z\over N}\,,\label{20b}
\end{alignat}
from which statements 1 and 2 made in the Introduction follow.
Indeed, let $z=z(t_i)>N/2$, then the probabilities that at 
time $t_{i-1}$ or $t_{i+1}$ the state was or will be $z-1$ 
is, in both cases, given by  ${z\over N}$. The probability 
for the state $z+1$ at time  $t_{i-1}$ or $t_{i+1}$ is ${N-z\over N}$.
Now, every change of state in the direction of the equilibrium 
distribution leads to an increase in entropy (see below). 
Hence the probability of having a higher entropy at 
$t_{i-1}$ or $t_{i+1}$ is ${z\over N-z}$ times that of having 
a lower entropy. If  $z=z(t_i)<N/2$ we have to use the inverse 
of that. 

\subsection{Boltzmann Entropy}
Boltzmann Entropy $S_B$ is a function $S_B:\Omega\rightarrow\reals$.
We stress that since $\Omega$ is defined only \emph{after} a choice 
of coarse graining (i.e. a choice of $\O_{\rm re}$) has been made, 
Boltzmann Entropy, too, must be understood as relative to that 
choice.\footnote{This apparently 
non-objective character of 
entropy is often complained about. But this criticism is based
on a misconception, since the term \emph{thermodynamical system}
is not defined without a choice for $\O_{\rm re}$. This is no 
different in phenomenological thermodynamics, where the choice 
of `work degrees of freedom', $\{y^i\}$, (the `relevant' or 
`controlled' degrees of freedom) is part of the definition of 
`system'. Only after they have been specified can one 
\emph{define} the one-form of heat, called $\omega$, 
as the difference between the differential of total energy, 
$dE$, and the one-form of reversible work, called 
$\alpha:=f_idy^i$; hence 
$\omega:=dE-\alpha$. Note that neither 
$\omega$ nor $\alpha$ are exact. In particular, $\omega\ne dQ$ for some 
function of state $Q$. In contrast to $E$, which is a function of \emph{states}, 
$\omega$ and $\alpha$ are each a  
function of \emph{processes}, which means that
given a curve $\gamma$ on the manifold of
(equilibrium) states, $\omega$ and $\alpha$ can be evaluated on (i.e. 
integrated along) $\gamma$. 
But it is meaningless to ask for the 
`value' of heat and work  on states. 
The value of heat associated to a 
process depends on the choice of 
$\alpha$, which in turn  depends on 
the choice of `relevant' $\{y^i\}$.
Roughly speaking, heat is the amount of 
energy not transmitted in the channels 
(degrees of freedom) controlled by the
$\{y^i\}$. This dependence of heat 
on the $\{y^i\}$ is directly 
inherited by entropy $S$, through 
$T\,dS=\omega$, where $T$ (temperature) 
and $S$ (entropy) are functions 
of state. They exist if and only if 
$\omega$ has an integrating factor 
(here $1/T$), which is the case if and 
only if $\omega\wedge d\omega=0$, or 
in differential-geometric terminology, if the kernel distribution of $\omega$ is integrable. This integrability is, in 
turn,  equivalent to the statement that 
in any neighbourhood of a given state 
there is another state that cannot 
be connected to the given one by a 
process (curve) on which the value of 
$\omega$ vanishes. To require that 
this latter be the case 
is just Carath{\'e}odory's \emph{principle of adiabatic inaccessibility} \citep{Caratheodory:1909}, which allows 
to deduce the existence of $S$ and whose
dependence on $\{y^i\}$ is now obvious. }
The value $S_B(z)$ in the macrostate $z$ is defined by 
$S_B(z):=\ln \mu_{\sss \Gamma}(\Gamma_z)$. For the urn model
this corresponds to the logarithm of microstates that 
correspond to the macrostate $z$. In what follows it will 
sometimes be more convenient to label the macrostate not by 
$z\in[0,N]$, but rather by a parameter $\sigma\in[-1,1]$
of range independent of $N$. Let the latter be defined by  
$z=\frac{N}{2}(1+\sigma)$. If we assume that $N,z,(N-z)\gg 1$
and approximate $\ln N!=N\ln N -N +O(\ln N)$ (Stirling formula),
we obtain the following expression for the Boltzmann entropy:
\begin{eqnarray}
S_B(z)&=&N\ln N-z\ln z-(N-z)\ln (N-z),
\label{21a}\\
&&\nonumber\\
S_B(\sigma)&=&-\frac{N}{2}\left[\ln\frac{1-\sigma^2}{4}
             +\sigma\ln\frac{1+\sigma}{1-\sigma}\right].
\label{21b}
\end{eqnarray}           
It obeys $S_B(\sigma)=S_B(-\sigma)=S_B(\vert\sigma\vert)$, 
which just corresponds to the invariance of the first expression 
under $z\mapsto N-z$. Considered as function of $\vert\sigma\vert$,
$S_B:[0,1]\rightarrow [\ln 2^N,0]$ is strictly monotonically 
decreasing. That $S_B(\sigma=1)=0$ is best seen in the limit 
$z\rightarrow N$ of (\ref{21a}). Despite Stirling's approximation 
this value is, in fact, exact, as one easily infers from the 
fact that $z=N$ just corresponds to a single microstate.
In contrast, the given value at $\sigma=0$ is only approximately 
valid.

\subsection{Consequences 1 and 2}
The quantitative form of Consequences~1 and 2 
are given by the solution to the following 
exercises: Let the state at time $t_i$ be 
$z=z(t_i)$. Calculate the conditioned 
probabilities for
\begin{itemize}
\itemsep=3.5pt
\item[(i)]
\hspace{5pt}
$z(t_i)$ being a local maximum,
\item[(ii)]
\hspace{5pt}
$z(t_i)$ being a local minimum,
\item[(iii)]
\hspace{5pt}
$z(t_i)$ lying on a segment of positive 
slope,
\item[(iv)]
\hspace{5pt}
$z(t_i)$ lying on a segment of negative slope.
\end{itemize}
Let the corresponding probabilities be $W_{\max}(z)$, $W_{\min}(z)$, 
$W_{\up}(z)$, and $W_{\down}(z)$, respectively. These are each given 
by the product of one past and one future conditioned probability. 
This being a result of the Markovian character of the dynamics, 
i.e. that for given  $(z,t_i)$ the dynamical evolution
$(z;t_i)\rightarrow (z\pm 1;t_{i+1})$ is independent of $z(t_{i-1})$.
Using (\ref{20a}-\ref{20b}) we obtain:   
\begin{alignat}{3}
&W_{\max}(z) &&\,=\, W_{\av}(z-1\vert z)W_{\ret}(z-1\vert z)
             &&\,=\, \left({z\over N}\right)^2,   \label{22a}\\
&W_{\min}(z) &&\,=\, W_{\av}(z+1\vert z)W_{\ret}(z+1\vert z)
             &&\,=\,\left(1-{z\over N}\right)^2, \label{22b}\\
&W_{\up}(z)  &&\,=\, W_{\av}(z-1\vert z)W_{\ret}(z+1\vert z)
             &&\,=\, {z\over N}\left(1-{z\over N}\right),\label{22c}\\
&W_{\down}(z)&&\,=\, W_{\av}(z+1\vert z)W_{\ret}(z-1\vert z)
             &&\,=\, {z\over N}\left(1-{z\over N}\right).\label{22d}
\end{alignat}
For $z/N>\frac{1}{2}$ ($z/N<\frac{1}{2}$) the probability 
$W_{\max}$ ($W_{\min}$) dominates the other ones. Expressed
in terms of $\sigma$ the ratios of probabilities are given 
by the simple expressions:
\begin{equation}
W_{\max}(\sigma):W_{\min}(\sigma):W_{\up}(\sigma):W_{\down}(\sigma)
={1+\sigma\over 1-\sigma}\,:\,{1-\sigma\over 1+\sigma}\,:\,1\,:\,1.
\label{23}
\end{equation}
In the limiting case of infinitely many $t_i$ we get that 
the state $z$ is $z^2/(N^2-z^2)=(1+\sigma)^2/2(1-\sigma)$
times more often a maximum than any other of the remaining 
three possibilities. 

We also note an expression for the expected recurrence time, 
$T(z)$, for the state $z$.\footnote{Note that we talk about 
recurrence in the space $\Omega$ of macrostates (`coarse 
grained' states), not in the space $\Gamma$ of microstates.}  
It is derived in \citet{Kac:1947} (there formula (66)). If the 
draws from the urns have constant time separation $\Delta t$
one has
\begin{equation}
T(z)={\Delta t\over W_{\ap}(z)},
\label{24}
\end{equation}
and hence a connection between mean recurrence time and
entropy:
\begin{equation}
S(z)=\ln\left[{2^N\Delta t\over T(z)}\right].
\label{25}
\end{equation}
\citet{Kac:1947} also shows the recurrence theorem,
which for discrete state spaces asserts the recurrence 
of each state with certainty. More precisely: let 
$W'(z';t_{i+n}\vert z;t_i)$ be the probability that 
for given state $z$ at time $t_i$ the state $z'$ occurs 
at time $t_{i+n}$ for the \emph{first} time after $t_i$
(this distinguishes $W'$ from $W$), then 
$\sum_{n=1}^{\infty} W'(z;t_{i+n}\vert z;t_i)=1$.
  
\subsection{Coarse grained Gibbs entropy and the H-theorem}
We recall that the Gibbs entropy $S_G$ lives on the space of 
probability distributions (i.e. normed measures) on $\Gamma$ 
and is hence independent of the choice of $\O_{\rm re}$.
In contrast, the coarse grained Gibbs entropy, $S_G^{cg}$,
lives on the probability distributions on $\Omega$,
$S_G^{cg}:\W\rightarrow\reals$, and therefore depends on
$\O_{\rm re}$. Since the former does serve, after all, as a 
$\O_{\rm re}$ independent definition of entropy (even though, 
thermodynamically speaking, not a very useful one), we 
distinguish the latter explicitly by the superscript `$cg$'.
If at all, it is $S_G^{cg}$ and not $S_G$ that thermodynamically 
can we be compared to $S_B$. The function $S_G^{cg}$ is given by   
\begin{equation}
S_G^{\rm cg}(W)=-\sum_{z=0}^N W(z)\cdot\ln\left[{W(z)\over W_{\g}(z)}\right].
\label{26}
\end{equation}
The structure of this expression is highlighted by means of the 
generalized $H$-theorem, which we explain below.%
\footnote{Usually this expression is called the 
\emph{relative} entropy [of $W$ relative to $W_{\g}$]. As [absolute] 
entropy of $W$ one then understands the expression 
$-\sum_zW(z)\ln W(z)$. The $H$-theorem would be valid for the 
latter only if the constant distribution (in our case $W(z)=1/(N+1)$)
is an equilibrium distribution, which is not true for the urn model.}
Since the two entropies  $S_B$ and $S_G^{cg}$ are defined on 
different spaces, $\Omega$ and $\W$, it is not immediately clear 
how to compare them. To do this, we would have to agree on what 
value of  $S_G^{cg}$ we should compare with $S_B(z)$, i.e. 
what argument $W\in\W$ should correspond to $z\in\Omega$. 
A natural candidate is the distribution centered at $z$, that is, 
$W(z')=\delta_z(z')$, which is 1 for $z'=z$ and zero otherwise. 
From (\ref{26}) we then obtain 
\begin{equation}
\label{eq:CompareEntropies}
S_G^{\rm cg}(\delta_z)=S_B(z)-N\ln 2\,.     
\end{equation}

Let us now turn to the generalized $H$-theorem. Let 
$\Phi:\reals\rightarrow\reals$ be a convex function. Then for any 
finite family $m:=\{x_1,\dots, x_n\}$ of not necessarily pairwise
distinct points in $\reals$ we have the following inequality  
$\Phi(\sum_i\alpha_ix_i)\leq\sum_i\alpha_i\Phi(x_i)\,\forall\,
\alpha_i\in\reals_{\geq 0}$ with $\sum_i\alpha_i=1$, where equality 
holds iff there is no index pair $i,j$, such that $x_i\not = x_j$ 
and $\alpha_i\cdot\alpha_j\not =0$. In the latter case the convex 
sum is called trivial. We now define a function 
$H:\W\times\W\rightarrow\reals$ through
\begin{equation}
H(W,W'):=\sum_{z=0}^NW'(z)\Phi\left[{W(z)\over W'(z)}\right].
\label{27}
\end{equation}
Consider a time evolution $W_i\mapsto W_{i+1}$,   
$W_{i+1}(z):=\sum_iW(z\vert z')W_i(z')$, where clearly 
$W(z\vert z')\geq 0$ and $\sum_zW(z\vert z')=1$. We also 
assume that no row of the matrix $W(z\vert z')$ just 
contains zeros (which would mean that the state labelled by the
corresponding row number is impossible to reach). We call such
time evolutions and the corresponding matrices \emph{non-degenerate}.
In what follows those distributions  $W\in\W$ for which 
$W(z)>0\,\forall z$, i.e. from the interior $\Wop\subset\W$,  
will play a special role. We call them \emph{generic}. 
The condition on $W(z\vert z')$ to be non-degenerate then 
ensures that the evolution leaves the set of generic distributions 
invariant. After these preparations we formulate      
\begin{theorem}[generalized H-theorem] 
Let $W'_i$ be generic and the time evolution non-degenerate; 
then $H(W_{i+1},W'_{i+1})\leq H(W_i,W'_i)$.
\end{theorem}

\begin{proof}
(Adaptation of the proof of theorem~3 in 
\citet{Kubo:1981} for the 
discrete case.) We define a new matrix 
$V(z,\vert z'):=[W'_{i+1}(z)]^{-1}W(z\vert z')W'_i(z')$, 
which generates the time evolution for $W_i(z)/W'_i(z)$ and obeys 
$\sum_{z'}V(z\vert z')=1$. It follows:
\begin{eqnarray}
H(W_{i+1},W'_{i+1})
&=&\sum_{z=0}^N W'_{i+1}(z)\,
\Phi\left[{W_{i+1}(z)\over\W'_{i+1}(z)}\right]          \label{28a}\\
&=&\sum_{z=1}^N W'_{i+1}(z)\,
\Phi\left[\sum_{z'=0}^N 
V(z\vert z'){W_i(z')\over W'_i(z')}\right]              \label{28b}\\
&\leq&\sum_{z'=0}^N\sum_{z=0}^N W'_{i+1}(z)V(z\vert z')\,
\Phi\left[{W_i(z')\over W'_i(z')}\right]                \label{28c}\\
&=&\sum_{z'=0}^N W'_i(z')\,
\Phi\left[\frac{W_i(z')}{W'_i(z')}\right]               \label{28c'}\\
&=&H(W_i,W'_i) \,.                                         \label{28d}
\end{eqnarray}
Equality in (\ref{28c}) holds, iff the convex sum in the 
square brackets of (\ref{28b}) is trivial.
\end{proof}

Picking a stationary distribution for $W'$, which in our case is the 
unique distribution $W_{\g}$, then $H$ is a function of just one 
argument which does not increase in time. Taking in addition the 
special convex function $\Phi(x)=x\ln(x)$, then we obtain with 
$S_G^{cg}:=-H$ the above mentioned entropy formula.  

Let from now on $\Phi$ be as just mentioned. Then we have, due to
$\ln(x)\geq 1-x^{-1}$, with equality iff $x=1$:
\begin{eqnarray}
H(W,W')&=&\sum_{z=0}^{N}W(z)\,\ln\left[{W(z)\over W'(z)}\right]
       \geq\sum_{z=0}^N(W(z)-W'(z))=0,\label{29'}\\
       &=&0\,\Leftrightarrow\, W(z)=W'(z)\quad\forall z.
\label{29}
\end{eqnarray}

Let us denote by a \emph{distance function} on a set $M$ any 
function $d:M\times M\rightarrow\reals_{\geq 0}$, such that 
$d(x,y)=d(y,x)$ and $d(x,y)=0\Leftrightarrow x=y$. (This is 
more general than a \emph{metric}, which in addition must satisfy 
the triangle inequality.) A map $\tau:M\rightarrow M$ is called 
non-expanding with respect to $d$, iff  
$d(\tau(x),\tau(y))\leq d(x,y)\,\forall x,y\in M$. We have 
\begin{theorem}
$D:\Wop\times\Wop\rightarrow\reals$, $D(W,W'):=H(W,W')+H(W',W)$ 
is a distance function with respect to which every proper 
non-degenerate time evolution is non-expanding.
\end{theorem}
\begin{proof}
Symmetry is clear and (\ref{29}) immediately implies 
$D(W,W')\geq 0$ with equality iff $W=W'$, as follows 
from the separate positivity of each summand. Likewise
(\ref{28d}) holds for each summand, so that no distance 
increases.
\end{proof}

\section{Thermodynamic limit and deterministic dynamics}
In this section we wish to show how to get a deterministic 
evolution for random variables in the limit $N\rightarrow\infty$.
To this end we first consider the discrete, future directed time 
evolution of the expectation value of the random variable $X(z)=z$.
We have 
\begin{eqnarray}
\Ev(X,t_{i+1})
&=&\sum_{z'=0}^Nz'W_{i+1}(z')
 =\sum_{z'=0}^N\sum_{z=0}^Nz'W_{\ret}(z'\vert z)W_i(z)\label{30}\\
&=&\sum_{z=0}^N\left[(z+1){N-z\over N}
             +(z-1){z\over N}\right]\, W_i(z) \nonumber\\
&=&1+\left(1-{2\over N}\right)\, \Ev(X,t_i).\label{31}
\end{eqnarray}
In the same way we get 
\begin{eqnarray}
\Ev(X^2,t_{i+1})
&=&\sum_{z=0}^N\left[(z+1)^2{N-z\over N}
  +(z-1){z\over N}\right]\, W_i(z) \nonumber\\
&=&1+2\Ev(X,t_i)+(1-4/N)\,\Ev(X^2,t_i)\,,\label{32}\\
&&\nonumber\\
\V(X,t_{i+1})
&=&\Ev(X^2,t_{i+1})-\Ev^2(X,t_{i+1})\nonumber\\
&=&(1-4/N)\, \V(X,t_i)
  +{4\over N}E(X,t_i)-{4\over N^2}E^2(X,t_i)\,.
\label{33}
\end{eqnarray}
By the evolution being `future directed' one means that 
$W_{\ret}$ and not $W{\av}$ are used in the evolution
equations, as explicitly shown in (\ref{30}). In this 
case one also speaks of `forward-directed evolution'.

In order to carry out the limit $N\rightarrow\infty$
we use the new random variable $\Sigma:\Omega\rightarrow\sigma$,
where $\sigma={2z\over N}-1$ as above; hence 
$X={N\over 2}(1+\Sigma)$. Simple replacement yields
\begin{eqnarray}
\Ev(\Sigma,t_{i+1})
&=&(1-2/N)\ \Ev(\Sigma,t_i)\,,\label{34}\\
\V(\Sigma,t_{i+1})
&=&(1-4/N)\ \V(\Sigma,t_i)
   +{4\over N^2}\left(1-\Ev^2(\Sigma,t_i)\right)\,.\label{35}
\end{eqnarray}
In order to have a seizable fraction of balls moved within 
a macroscopic time span $\tau$, we have to appropriately 
decrease the time steps $\Delta t:=t_{i+1}-t_i$ with growing 
$N$, e.g. like  $\Delta t={2\over N}\tau$, where $\tau$ is some 
positive real constant. Its meaning is to be the time span, in 
which $N/2$ balls change urns. Now we can take the limit 
$N\rightarrow\infty$ of (\ref{34}) and (\ref{35}),
\begin{alignat}{4}
& {d\over dt}E(\Sigma,t)
&&\,=\,-{1\over\tau}\Ev(\Sigma,t)
&&\,\Longrightarrow\, \Ev(\Sigma,t)
&&\,=\,E_0\exp\left({-(t-t_1)\over\tau}\right), \label{36}\\
& {d\over dt}\V(\Sigma,t)
&&\,=\,-{2\over\tau}\V(\Sigma,t)
&&\,\Longrightarrow\, \V(\Sigma,t)
&&\,=\,V_0\exp\left({-2(t-t_2)\over\tau}\right), \label{37}
\end{alignat}
where $E_0,V_0,t_1,t_2$ are independent constants. 
These equations tell us, that 1)~the expectation value 
approaches the equilibrium value  $\Sigma=0$ exponentially 
fast in the future, and 2)~it does so with exponentially 
decaying standard deviation. The half mean time 
of both quantities is the time for $N/2$ draws. 

According to the discussions in previous sections it is now 
clear, that in case of equilibrium identical formulae would have 
emerged if $W_{\av}$ instead of $W_{\ret}$ had been used, for 
then $W_{\av}=W_{\ret}$. Most importantly to note is, that the 
backward evolution is \emph{not} obtained by taking the forward 
evolution and replacing in it $t\mapsto -t$. The origin of this 
difference is the fact already emphasized before (following 
Theorem\,2), that $W_{\av}(z;z')$ is not the inverse matrix 
to $W_{\ret}(z;z')$, but rather the matrix computed according to 
Bayes' rule.    

\section{Appendix}
In this Appendix we collect some elementary notions of 
probability theory, adapted to our specific example. 

The space of elementary events\footnote{`Elementary' is merely 
to be understood as mathematical standard terminology, not in any 
physical sense. For example, in the urn model, $\Omega$ is obtained 
after coarse graining from the space of physically `elementary'
events.} is $\Omega=\{0,1,\dots,N\}$. By
\begin{eqnarray}
\X:&=&\bigl\{X:\Omega\rightarrow\reals\bigr\}\,,
\label{A1}\\
\W:&=&\bigl\{W:\Omega\rightarrow\reals_{\geq 0}\mid 
         \sum_{z\in\Omega}W(z)=1\bigr\}\,,
\label{A2}
\end{eqnarray}
we denote the sets of random variables and probability 
distributions respectively, where $\W\subset\X$. 
The map $\X\rightarrow\reals^{N+1}$, 
$X\mapsto (X(0),X(1),\cdots,X(N))$ defines a bijection 
which allows us to identify $\X$ with $\reals^{N+1}$.
This identifies $\W$ with the $N$-simplex 
\begin{equation}
\label{eq:N-Simplex}
\Delta^N:=\bigl\{(W(0),\cdots,W(N))\in\reals^{N+1}\mid W(z)\geq 0,\,
\sum_{z}W(z)=1\bigr\} \subset\reals^{N+1}\,.
\end{equation}
Its boundary, $\partial\Delta^N$, is the union of all  
$(N-K)$-simplices:
\begin{equation}
\label{eq:N-SimplexBoundary}
\Delta^{i_1\cdots i_K}:=\bigl\{(W(0),\dots,W(N))\in\Delta^N\mid 
0=W(i_1)=\cdots=W(i_K)\bigr\} \,,
\end{equation}
for all $K$. Its interior is $\Wop:=\W-\partial\W$, so that 
$W\in\Wop\Leftrightarrow W(z)\not =0\forall z$.

Expectation value $\Ev$, variance $\V$, and standard deviation $\S$  
are functions $\X\times\W\rightarrow\reals$, defined as follows:
\begin{alignat}{2}
&  \Ev:\X\times\W\rightarrow\reals,\quad
&& \Ev(X,W):=\sum_{z\in\Omega}X(z)W(z) \,,          
   \label{A3}\\
&  \V:\X\times\W\rightarrow\reals_{\geq 0},\quad
&& 
\V(X,W):=\Ev((X-\langle X\rangle)^2,W)=\Ev(X^2,W)-\Ev^2(X,W)\,,
\label{A4}\\
&  \S:\X\times\W\rightarrow\reals_{\geq 0},\quad
&& \S(X,W):=\sqrt{\V(X,W)}\,,
\label{A5}
\end{alignat}  
where in (\ref{A4}) $\langle X\rangle$ simply denotes the 
constant function $\langle X\rangle:z\mapsto \Ev(X,W)$, and 
$\Ev^2(X,W):=[\Ev(X,W)]^2$. In the main text we also write
$\Ev(X,s)$ if the symbol $s$ uniquely labels a point in $\W$,
like $s=\text{ap}$ for the a priori distribution (\ref{1}),
or $E(X,t_i)$ for the distribution $W_i$ at time $t_i$.

\bibliography{Entropy.bib}
\bibliographystyle{plainnat}
\end{document}